\documentclass[twocolumn,showpacs,preprintnumbers,amsmath,amssymb,prl]{revtex4}
\usepackage{graphicx,dcolumn,bm,eucal,hyperref}

\IfFileExists{srcltx.sty}{\usepackage[active]{srcltx}}

\newcommand{\cll}{$\chi$LL\ \unskip}

\begin{document}

\title{Microscopic construction of the chiral Luttinger liquid theory of the
  quantum Hall edge.}

\author{A. Boyarsky$^{1},$ Vadim~V.~Cheianov$^{2}$ and O. Ruchayskiy$^{3}$}
\affiliation{$^1$EPFL, BSP-Dorigny, CH-1015, Lausanne,
  Switzerland\\$^2$NORDITA, Blegdamsvej 17, Copenhagen {\O}, DK 2100,
  Denmark\\$^3$IHES, Bures-sur-Yvette, F-91440, France }

\date{\today}

\begin{abstract}
  We give a microscopic derivation of the chiral Luttinger liquid theory ($\chi$LL)
  for
  the Laughlin states. Starting from the wave function describing an
  arbitrary incompressibly deformed Laughlin state (IDLS) we quantize these
  deformations. In this
  way we obtain the low-energy projections of local microscopic operators and derive  
  the quantum field theory of edge excitations directly from quantum mechanics of
  electrons. This shows that to describe experimental and numeric deviations
  from  $\chi$LL one needs to go beyond Laughlin's approximation.
  We show that in the large N limit the IDLS is described by
  the dispersionless Toda hierarchy.
\end{abstract}
\maketitle %

The fractional quantum Hall effect is a manifestation of strong correlations
in the two dimensional electron gas in a strong magnetic field
\cite{Prange-1990}. At certain filling fractions $\nu = 2 \pi l^2 \rho,$ where
$l$ is the magnetic length and $\rho$ is the electron density, the electrons
``condense'' into an incompressible fluid, which leads to the quantization of
the Hall conductance. While the spectrum of bulk excitations of the
incompressible Hall state is gapful, the boundary of such a state exhibits
rich low-energy physics associated with gapless edge modes.  In 1990, based on
elegant arguments of locality, chirality and gauge invariance, X.-G. Wen
proposed an effective field theory for the $\nu=1/(2p+1)$ quantum Hall edge
\cite{Wen-1990}. Later on, this theory was generalized to other filling
factors at which the incompressible quantum Hall states
occur~\cite{OtherFillingFactors,Frohlich}.  The construction given
in~\cite{Wen-1990}, is usually referred to as the chiral Luttinger liquid
($\chi$LL).  The $\chi$LL theory predicts that once the Hall conductance of
the incompressible bulk state is known, the low-energy properties of the edge
become independent of the details of electron-electron interactions and the
confining potential, the only non-universal parameter being the propagation
velocity of the edge excitations.  Probably, the most intriguing prediction of
the $\chi$LL theory is the universal power law scaling of local operators and
in particular the electron spectral weight at the edge.

Early attempts to put the $\chi$LL theory to the test were quite successful.
The compressible edge modes were observed in a number of experiments.  The low
energy excitation spectrum studied by exact diagonalization of small systems
at $\nu=1$ \cite{Stone-1992} and at $\nu=1/3$ \cite{Palacios-1995} was found
to be consistend with the predictions of $\chi$LL. The ground state momentum
distribution function was tested numerically for the Laughlin state in various
geometries \cite{Rezayi-1994,Datta-1996} and was found to be in agreement with
the predictions of $\chi$LL. Recently, however, both experimental
\cite{Hilke-2001} and numerical \cite{Goldman-2001,Tsiper-2001,Jain-2001}
evidences emerged that the scaling properties of the observables at the
quantum Hall edge may actually be non-universal. Moreover, in a recent
work~\cite{Zulicke-2003} very interesting numerical results were reported,
indicating that one of the key assumptions of the $\chi$LL theory, namely the
assumption that the electron creation (annihilation) operators preserve their
fermionic statistics in the low-energy effective theory, may be violated for
some quantum Hall states.  This results question the universality of the
$\chi$LL theory and motivate a search for microscopic approaches to the
description of the quantum Hall edge.  This task can be pursued in two
different ways.  One is to use the composite fermion picture
\cite{Chklovskii-1995}.  The problem of this approach is that at some point it
has to rely on the mean-field approximation, which, if valid, is still
extremely hard to justify.  Another way is to use trial wave functions to
build the low energy sector of the Hilbert space and to construct the low
energy projections of local operators explicitly. Such a construction was
discussed in~\cite{Filippo-1996,Zulicke-1999}, where the low-energy
projections of the edge current operators were studied. In~\cite{Zulicke-1999}
a local operator for the periphery deformation accommodating a unit charge was
also proposed. In this paper we use a similar philosophy to give a direct
(rather than phenomenological) derivation of the \cll\ theory, starting from
quantum mechanical many-particle wave functions.  Here it will be done for the
Laughlin state. From our construction it becomes clear that one needs to 
go beyond Laughlin's approximation to understand the experimental and numerical
deviations from the $\chi$LL theory. At the same time, being microscopic in nature,
our approach introduces a framework for field-theoretical treatment of 
corrections to the Laughlin's approximate ansatz.

One way to view the $\chi$LL theory of
the quantum Hall edge is to consider it as a dynamical description of the
low-energy projections of the microscopic operators.  Formally, for a given
operator $A$ its low energy projection, which we will denote as $\hat A,$ is
defined as
\begin{equation}
\hat A= \mathcal P A \mathcal P, \qquad
\mathcal P =\sum_{\vert \nu\rangle \in \mathcal H} \vert\nu\rangle \langle  \nu \vert,
\label{lepdef}
\end{equation}
where $\mathcal H$ is the Hilbert space of such low-energy excitations.
The $\chi$LL construction~\cite{Wen-1990,OtherFillingFactors} relies on some
assumptions about the low energy projection of microscopic operators. Those
include: chirality; current algebra; charge and statistic of the projected
fermion operators.  Neither reliability nor limitations of any of these
assumptions can be understood without the understanding of the properties of
the low energy projection. In the following paragraphs we use trial wave
functions for a consistent microscopic construction of the projected currents
and the projected local electron operator and check their physical properties
for the Laughlin states $\nu=1/(2p+1).$

We base our considerations on the $N$-particle Laughlin's wave function
\cite{Laughlin-1983}
\begin{equation}
\langle z_1,\dots , z_N\vert 0\rangle=\frac{1}{\sqrt N!}
\prod_{i<j}^{N}(z_j-z_i)^\frac{1}{\nu} \prod_{j=1}^{N}e^{-\frac{1}{2 h \nu }
  z_j\bar z_j} 
\label{Laughlin}
\end{equation}
which is known to describe the ground state $\vert 0\rangle$ of a disk-shaped
droplet of the quantum Hall fluid with remarkable accuracy
\cite{Haldane-1983}.  The parameter $h=2 l^2/ \nu $ was introduced in
Eq.~\eqref{Laughlin} to shorten notation. Laughlin pointed out
\cite{Prange-1990} that by the following deformation of the ground state:
\begin{equation}
\langle z_1, \dots, z_N\vert t_k\rangle=\langle z_1, \dots, z_N\vert 0\rangle
\prod_{j=1}^{N} e^{\frac{1}{h \nu}\omega(z_j)},
\label{tstatedef}
\end{equation}
the shape of the quantum Hall droplet can be changed in such a way as to
preserve the electron density and the area of the droplet. In
Eq.~(\ref{tstatedef})
\begin{equation}
\omega(z)\equiv\sum_{k>0} t_k z^k
\label{omegadef}
\end{equation}
is an entire function of $z$ depending on the complex parameters $t_k$. Below,
we will often refer to the state Eq.~\eqref{tstatedef} as to the
\emph{incompressibly deformed Laughlin state} (IDLS).  The geometry of the
IDLS for $\nu =1$ was investigated in the recent work~\cite{wz-qhe}. It was
explicitly shown that in the large $N$ limit the charge density, defined as $
\rho(\xi,\bar\xi) =\mathrm{const} \int d\mu \vert \langle z_1,
\dots,z_{N-1},\xi\vert t_k\rangle \vert^{2},$
where $d\mu\equiv d^2 z_1\dots d^2 z_N,$ is given by the two-dimensional
step-function, with the support in the region, bounded by the curve $\gamma$
whose harmonic moments are equal to $t_k$ from Eq.~(\ref{omegadef}), i.e.
\begin{equation}
  \label{tk}
 \oint_\gamma
\frac{dz}{2\pi k}\frac{\bar z}{z^k} =-\frac{1}{\pi k} \int_{D_{-}} z^{-k} d^2z= t_k
\end{equation}
(here $D_{-}$ stands for the \emph{exterior} of the droplet with the boundary
$\gamma$, $D_+=\mathbb{C}/D_-$ is an \emph{interior} of the domain).  In fact,
it is also true for $\nu<1$~\cite{to_appear}.  The key result is that the norm
of the IDLS
\begin{equation}
\tau^\nu_N(t_k, \bar t_k)=
\int d\mu \vert \langle z_1, \dots,z_N\vert t_k\rangle \vert^{2}
\label{taudef}
\end{equation}
is given in the large $N$ limit ($N\to\infty$, $h\to0$, $t_0 = Nh$ is being
kept fixed) by
\begin{equation}
\tau_N^\nu(t_k,\bar t_k)\to\tau^{\nu}(t_0, t_k,\bar t_k)= e^{{h^{-2}\nu^{-1}}
 F(t_0, t_k, \bar t_k)},
 \label{Fractau}
\end{equation}
where $F$ is the logarithm of the dispersionless tau-function of 2D Toda
lattice hierarchy~\cite{Kostov-2000}. Note, that the function $F$ is \emph{the
  same} for both $\nu=1$ and $\nu<1$.  
It will be important that
for $t_k,\bar t_k=0$ (from now on we choose $t_0 =1$ unless explicitly stated
otherwise)
\begin{equation}
\frac{\partial F}{\partial t_p}=
\frac{\partial F}{\partial\bar t_p}=0
\quad\mbox{and}\quad
\frac{\partial^2 F}{\partial \bar t_q\partial t_p}
=q\delta_{q,p}
\label{F2d}
\end{equation}

Next, we calculate the energy of the IDLS.
Note that the incompressible deformation  does not influence
the short-range exchange-correlation part of the interaction energy since the
local structure of the wave function is unaffected. Thus the only change of
the energy comes from the Hartree term:
$E(t_k,\bar t_k)=\int\hskip -.5em\int_{D_+} \hskip -.25em d^2 z d^2 w\,
V(\vert z- w\vert)+\int_{D_+} \hskip -.25em d^2 z\, U(\vert z\vert)$.
Here $V(r)$ is the electron-electron repulsion potential and $U(r)$ is the axially symmetric
confining potential, which keeps the Laughlin state from breaking apart. %For simplicity we
The Hilbert space $\mathcal H_N$ of the low-energy excitations of the
$N$-particle incompressible Laughlin droplet is defined as that of IDLS with
the energy less than some cutoff~$\Lambda$. For such states one has
\begin{equation}
E(t_k,\bar t_k)= E_0+ \sum_{k>0} s(k) k^2  t_k \bar t_k,\quad \sum_{k>0} s(k)
t_k \bar t_k <\Lambda 
\label{E(t)}
\end{equation}
where $s(k)$ is given by the following expression:
\begin{align}
%  s(k)&=\frac{\pi \nu}{ h^2} U_0 k^2 +\int_{\raisebox{-.3em}{$\scriptstyle
%      0$}}^{\raisebox{.2em}{$\scriptstyle 2 \pi$}} \hskip -1em {d \theta}\,
%  R^2 V(2 R \vert \sin \theta \vert) \left[\cos k \theta-\cos\theta \right]
  s(k)&=\frac{\pi}{ h^2} U_0 k^2 +\int_{\raisebox{-.3em}{$\scriptstyle
      0$}}^{\raisebox{.2em}{$\scriptstyle 2 \pi$}} \hskip -1em {d \theta}\,
  V(2 \vert \sin \theta \vert) \left[\cos k \theta-\cos\theta \right]
\label{sk}
\end{align}
For small $k$ Eq.~\eqref{sk} gives $s(k)=s_0 k^2$ where $s_0$ is a \emph{Fermi
  velocity}.
%where $s_0$ is a non-universal $k$-independent constant.  

The projection of an arbitrary state $\Phi(z_1, \dots , z_N)$ in $\mathcal
H_N$ onto the incompressibly deformed Laughlin state $\Psi(z_1, \dots, z_N\vert t_k)$
defined in Eq.~\eqref{tstatedef}
\begin{equation}
\Phi(\bar t_k)=\int d\mu\,\bar \Psi(\bar z_1, \dots, \bar z_N\vert \bar t_k)
\Phi(z_1, \dots , z_N)
\label{bartrep}
\end{equation}
defines the \emph{$\bar t$-rep\-re\-sen\-tation} of the state $\Phi.$ As we will see, it is
particularly convenient for building the low-energy projections of the
microscopic operators.
It is important to know how the inner product of two states in the $\bar t$
representation is calculated. Introduce the following basis of states
\begin{align}
\vert \vec n \rangle=\left.
\frac{1}{\sqrt{\tau^{\nu}} }\prod_{k=1}^\infty
 \left(\frac{\partial}{\partial t_k}\right)^{n_k} \Psi_N(z_1\dots z_N\vert
 t_k)\right|_{t_k=0}, 
 \label{basis}
\end{align}
where $\vec n=(n_1, \dots, n_p, \dots )$ is a vector with non-negative integer
valued components. Here $\tau^{\nu}$ in the denominator is the shorthand for
$\tau^{\nu}(t_0, 0,0).$ Since any entire function of $t_k$ can be expanded in
Taylor series, the states Eq.~\eqref{basis} form an (over)complete basis in
the subspace $\mathcal H_N.$ The Gram matrix of the basis Eq.~\eqref{basis} is
calculated as
\begin{equation}
\langle \vec n\vert  \vec m\rangle=\frac{1}{\tau^{\nu}} \prod_{k,p\ge 1}
\left.\frac{\partial^{m_k+n_p}\tau^{\nu}(t_k,\bar t_k) }
{\partial t_k^{m_k}\partial \bar t_p^{n_p}}\right\vert_{t_k, \bar t_k=0}
\end{equation}
By using the asymptotic formulas Eqs.~\eqref{Fractau},~\eqref{F2d} one finds
that to the leading order in $h$ basis~\eqref{basis} is
orthogonal~\footnote{States $|\vec n\rangle$ are linearly independent only for
  $\sum_k k n_k < N$. If this condition does not hold, neither does
  Eq.~(\ref{sprod}). From Eq.~(\ref{eigenval}) it follows that this condition
  coincides with that of in Eq.~(\ref{E(t)}) for $\Lambda\sim \nu h^2 s_0 N$.}
\begin{equation}
\langle \vec n\vert  \vec m\rangle= C_{\vec n}\:\delta_{\vec m, \vec
  n}\:,\;\mbox{~~~where~~~}\:
C_{\vec n}= \prod_{k\ge1}  \left(\textstyle\frac{k}{\nu h^2}\right)^{n_k} n_k!
\label{sprod}
\end{equation}
From definitions Eqs.~\eqref{bartrep} and \eqref{basis} it follows that the
projection of an arbitrary state $\Phi$ in $\bar t$-representation onto the
state $\vert \vec n\rangle$ is given by
\begin{equation}
\langle \vec n \vert \Phi \rangle =
\frac{1}{\sqrt {\tau^{\nu}}}
\prod_{k\ge 1}\left(\frac{\partial }{\vphantom{N^2}\partial \bar t_k}\right)^{n_k}
\Phi( \bar  t_k)
\label{projstate}
\end{equation}
Therefore, for two states $\Phi_a$ and $\Phi_b$ we have
\begin{equation}
\langle \Phi_a\vert \Phi_b\rangle=\frac{1}{\tau^{\nu}}
\sum_{\vec n}\prod_{k\ge1}
\left(\frac{\partial^2}{\partial t_k \partial \bar t_k} \right)^{n_k}
\frac{\bar \Phi_a(t_p) \Phi_b(\bar t_p)}{\langle \vec n \vert \vec n \rangle}
\label{derivrep}\end{equation}
Using Eq.~\eqref{sprod} one can see that the following compact
representation of Eq.~\eqref{derivrep} exists
\begin{equation}
\langle \Phi_a\vert \Phi_b\rangle=
\left.\bar
  \Phi_a\left(\textstyle\frac{h^2\nu}k\frac{\partial}{\vphantom{N^2}\partial
      \bar t_k}\right)\Phi_b(\bar t_k) \right|_{\bar t_k =0}
\label{innerint}
\end{equation}
The algebra of operators in $\bar t$-representation is generated by the
elements
\begin{equation}
\hat a_{-k}=  \sqrt{\frac{ k}{\nu h^2}}\:\bar t_k, \quad
\hat a_{k} ={\sqrt {\frac{\nu h^2}{ k} }}\frac{\partial}{\partial \bar t_k}, \quad
[\hat a_{k},\, \hat a_{-k}]=1
\label{adef}
\end{equation}
(for $k>0$).  From the inner product \eqref{innerint} one finds
\begin{equation}
\hat a_{-k}=\hat a_k^\dagger
\label{adag}
\end{equation}

The $\bar t$ representation has a natural extension to the low energy Fock space
$\mathcal H = \mathcal H_N\oplus 
\mathcal H_{N-1}\oplus\mathcal H_{N+1}\oplus \dots$
In this space $t_0=h N$ becomes an operator rather than a parameter. In the large 
$N$ limit one can neglect the constraint $t_0>0$ and realize the particle number 
operator $\hat t_0$ as the angular momentum operator conjugate to an
auxiliary angle variable $\varphi$ satisfying $\varphi =\varphi +2 \pi:$
\begin{equation}
\hat t_0=- i h \frac{\partial}{\partial \varphi}, \quad [\hat t_0, e^{\pm i \varphi}]=
\pm h e^{ \pm i \varphi}
\label{angular}
\end{equation}
Consider some $N$-particle IDLS $\vert t_0,t_k\rangle$, where $t_0=h N$
explicitly shows the number of particles.  This is an eigenstate of $\hat
t_0$. Any state $\vert \Phi \rangle$ where the number of particles is not
defined can be projected on $\vert t_0,t_k\rangle.$ This projection defines
the $\bar t$ representation in $\mathcal H$: $\Phi(t_0, \bar t_k)=\langle
t_0,t_k \vert \Phi \rangle$.  For example, the $\bar t$ representation of a
$N$-particle IDLS characterized by the harmonic moments $t_k^{\prime}$ will be
given by
\begin{equation}
\langle t_0, t_k \vert t_0', t^{\prime}_k\rangle=
\tau^{\nu}(t_0, t_k^{\prime} , \bar t_k)\delta_{t_0, t_0'}.
\label{idls}
\end{equation}
In the following we will use the explicit action of the phase operators $e^{i
  \varphi}$ in the $\bar t$ representation:
\begin{equation}
e^{i \varphi} \Phi(t_0, \bar t_k)= \Phi(t_0-h, \bar t_k).
\label{phaseaction}
\end{equation}
This formula completes the description of the $\bar t$-rep\-re\-sentation of
the space $\mathcal H$ of low-energy excitations.

It is easy to see that for positive $m$ ~$\hat a_m$ is proportional to the
Fourier component $\hat\rho_m$ of the edge density
operator~\cite{Wen-1990,Zulicke-1999} defined as
$\rho_m=(z_1^m+\dots+z_N^m)/\sqrt{2 \pi}$.  Namely from
Eqs.~\eqref{adef},~\eqref{tstatedef} and~\eqref{omegadef} one finds
\begin{equation}
  \label{charge}
\hat \rho_m=\hat a_m\sqrt{\frac{\nu
  |m|}{2\pi}}\quad\mbox{and}\quad  [\hat \rho_m,\,\hat \rho_{n}] = \frac{\nu\,
m}{2\pi} \delta_{m+n,0} 
\end{equation}
(By virtue of Eq.~\eqref{adag}, Eq.~\eqref{charge} is true for both positive
and negative $m$).  Thus we have derived microscopically the central charge of
the current algebra of the edge density operators, obtained in~\cite{Wen-1990}
from the gauge invariance arguments.

Consider now the microscopic fermion field $\psi(\zeta),$ which annihilates an
electron at the point $\zeta$. In $z$ representation it is defined as
\begin{equation}
  [\psi(\zeta)\Psi](z_1,
  \dots, z_N)= \sqrt{ N+1} \Psi(z_1,\dots,z_N, \zeta)
  \label{fermion}
\end{equation}
We will find the action of the conjugate (creation) operator $\psi^\dagger$ on
the $\bar t$ representation Eq.~(\ref{bartrep}) of the state
$|t_0',t_k'\rangle$, given by Eq.~(\ref{idls}).  The $\bar t$ representation
of the state $\psi^\dagger(\bar z)|t_0',t_k'\rangle$ is given by
\begin{equation}
\hat \psi^\dagger(\bar\zeta) \tau^{\nu}(t_0, t_k',
\bar t_k)\delta_{t_0,t'_0}=
\langle t_0, t_k \vert \psi^\dagger(\bar\zeta) \vert t'_0, t_k' \rangle.
\label{Amel}
\end{equation}
where $\hat\psi^\dagger$ on the l.h.s. is by definition the $\bar t$
representation of the low energy projection of the fermion operator. We start
from calculating the matrix element
\begin{eqnarray}
  &&\langle t_0,t_k \vert \psi^{\dagger}(\bar \zeta) \vert t_0',t_k\rangle=\\
  &&e^{-\frac{|\zeta|^2-2\bar \omega(\bar \zeta)}{2h\nu}}
  \delta_{t_0-h,t_0'}\int d \mu \prod_{j=1}^{N} (\bar \zeta-\bar
  z_j)^{\frac{1}{\nu}}
  \vert \langle z_1 \dots z_N\vert t_k\rangle \vert^{2}\nonumber\\
  &&= \bar\zeta^{\frac{t_0-h}{h\nu}}e^{-\frac{|\zeta|^2-2\bar \omega(\bar
      \zeta)}{2h\nu}}
  \delta_{t_0-h,t_0'}\times\tau^{\nu}\left(t_0-h, t_k, \bar
    t_k-\textstyle\frac{h}{k \bar \zeta^k}\right)\nonumber
 \label{psimel}
\end{eqnarray}
where $t_0=h (N+1) $. %  Here we used Eqs.~\eqref{Laughlin} and\eqref{tstatedef}. 
The last line in Eq.~\eqref{psimel} was obtained using the definitions
Eqs.~\eqref{taudef} and~\eqref{Fractau}, exponentiating $(\bar \zeta-\bar
z_j)^{\frac{1}{\nu}}$ and expanding the logarithm. Because of the fact that
$\tau^\nu(t_0,t_k,\bar t_k)$ is a function, analytic in both sets $t_k$ and
$\bar t_k$ independently, Eq.~\eqref{psimel} actually gives the r.h.s. of
Eq.~(\ref{Amel}) for arbitrary $t_k'$. This defines the action of the operator
$\hat \psi^\dagger$ in the $\bar t$ representation. Indeed, comparing
Eqs.~\eqref{Amel},~\eqref{psimel} and~\eqref{phaseaction} we get
\begin{equation}
\hat \psi^{\dagger}(\bar \zeta) =  e^{-\frac{1}{2h\nu}\vert \zeta\vert^2}
\,e^{ i \varphi} 
 : e^{\hat \phi(\bar \zeta)}:
\label{fermidag}
\end{equation}
where
\begin{equation}
\hat \phi(\bar \zeta)= \frac{\hat t_0}{h \nu}\ln(\bar \zeta)+
\sum_{k\ge1} \left[
\frac{1}{\sqrt{ \nu k} } \, \bar \zeta^k \hat a_{k}^{\dagger}
-
\frac{1}{\sqrt{\nu k}} \, \bar \zeta^{-k} \hat a_{k}   \right]
\label{phidef}
\end{equation}
The normal ordering $::$ is understood in the standard sense $:\exp(\hat
\phi):\:=\exp(\hat \phi_0)\exp(\hat \phi_{+}) \exp(\hat \phi_{-}),$ where
$\hat \phi_{+}$ is the positive frequency part of $\hat \phi,$ only containing
creation operators $\hat a_k^{\dagger},$ $\hat\phi_{-}$ is the negative
frequency part and $\hat\phi_0$ is the zero mode.  The field $\hat\phi$
satisfies the obvious property
%\begin{equation}
$
[\hat\phi(\bar \zeta)]^{\dagger}=-\hat\phi(1/\zeta).
\label{phiconj}
$
%\end{equation}
Therefore the electron annihilation operator is
\begin{equation}
\hat\psi(\zeta)= e^{-\frac{1}{2 h\nu}\vert \zeta\vert^2}
: e^{-\hat \phi(1/\zeta)}:e^{-i \varphi}.
\label{ferminedag}
\end{equation}
This completes the construction of the projected fermion operators.

Notice that since $|t_0',t_k\rangle$ span the whole Hilbert space $\mathcal
H$, we have defined the action of the operator $\hat\psi^\dagger(\bar\zeta)$
on any state in $\mathcal H$. In this way one can in principle find $\bar t$
representation of any microscopic operator.

Derive, for instance, the $\bar t$ representation of the Hamiltonian. By
virtue of Eq.~\eqref{E(t)} and with the help of the the relations
Eq.~\eqref{Fractau},~\eqref{F2d} one has
\begin{equation}
\langle  t_k \vert (H-E_0) \vert t_k\rangle={s_0 h^2 \nu }
\sum_{k\ge 1} k \bar t_k \frac{\partial}{\partial \bar t_k}
\tau^{\nu}(t_k, \bar t_k)
\end{equation}
According to the given above prescription the projected Hamiltonian is
immediately identified as
\begin{equation}
\hat H- E_0={s_0 h^2\nu} \sum_{k\ge1} k  \hat a_{-k} \hat a_{k}
\label{PHP}
\end{equation}
Note, that for the $\bar t$-representation of the vacuum (Laughlin) state,
$\tau^{\nu}(t_0, 0, \bar t_k)=\mathrm{const}$ we have
\begin{equation}
\hat a_{k}\vert 0\rangle =0, \quad k>0,
\label{vacanih}
\end{equation}
therefore the operator Eq.~\eqref{PHP} is normal ordered.  Notice that
Hamiltonian Eq.~(\ref{PHP}) coincides with the Virasoro generator $L_0$ of the
theory of free chiral scalar $\phi$.  From Eq.~\eqref{PHP} one can see that
states $\vert \vec n \rangle$ are indeed the eigenstates of the projected
Hamiltonian
\begin{equation}
(\hat H-E_0) \vert \vec n \rangle ={s_0 h^2\nu} \sum_{k\ge0} k n_k
\vert \vec n \rangle
\label{eigenval}
\end{equation}

The positive- and the negative-frequency parts
of $\hat \phi$ satisfy the following commutation relations
\begin{equation}
[\hat \phi_{-}(z), \hat \phi_{+}(w)]=-\sum_{k} \frac{1}{\nu k} \left(\frac{w}{z}\right)^k=
\frac{1}{\nu}\ln\left(1-\frac{w}{z}\right)
\label{commutatorpm}
\end{equation}
The correlation functions of the projected fermion operators can be obtained
in a standard way from Eqs.~\eqref{fermidag}, \eqref{ferminedag},
\eqref{vacanih}, \eqref{commutatorpm} and with the help of the formula
\begin{equation}
:e^{\hat \phi(\bar z)}: :e^{-\hat \phi(w) }:=:e^{\hat \phi(\bar z)-\hat \phi(w) }:
e^{-[\hat \phi_{-}(\bar z), \hat \phi_{+}(w)]}.
\end{equation}
For large enough distances ($|z-w|> h^2 s_0/\Lambda $) one has
\begin{equation}
\langle 0\vert  \psi^{\dagger}(\bar z) \psi(w) \vert 0\rangle=
e^{-\frac{1}{2h\nu}(\vert w\vert^2+\vert z\vert^2)}
\frac{(w \bar z)^{\frac{N}{\nu} } }{(1-\frac{1}{ w \bar z})^{1/\nu}}.
\label{psicoranswer}
\end{equation}
For $\vert z\vert =|w|=1$ this formula is very similar to Wen's result
\cite{Wen-1990}.  The difference between the correlation function proposed by
Wen and Eq.~\eqref{psicoranswer} is in the boundary conditions. While Wen's
correlation function corresponds to antiperiodic boundary conditions for the
fermion field $\psi(\theta),$ the microscopically derived correlation function
is periodic in $\theta$, which is more natural.

To calculate the time-dependent correlation function, we use the Hamiltonian
Eq.~\eqref{PHP}. One can easily see that
\begin{equation}
e^{i \hat H t} \hat \phi(z) e^{-i\hat H t}=\hat \phi\left(z e^{i \omega t}\right )
\end{equation}
where $\omega=h^2 \nu s_0$ (this formula is the microscopically derived
chirality condition for the field $\hat \phi$).  Thus
\begin{equation}
\langle 0\vert  \psi^{\dagger}(\bar z,t) \psi(w) \vert 0\rangle=
\langle 0\vert  \psi^{\dagger}(\bar z e^{-i \omega t}) \psi(w) \vert 0\rangle.
\end{equation}
This result also agrees with Wen's theory.
Finally, using the Campbell-Hausdorff formula and Eq.~\eqref{phaseaction}
for the product of two fields
Eq.~\eqref{ferminedag} one finds
\begin{equation}
\hat \psi(z)\hat \psi(w)=\left(\frac{w}{z}
\right)^{\frac{1}{\nu}}e^{[\hat \phi(1/z), \hat \phi(1/w)]}
\hat \psi(w)\hat \psi(z).
\end{equation}
This is formula is correct for large enough separations $\vert z- w\vert $.
Calculating the commutator in the exponential with the help of
Eq.~\eqref{commutatorpm} one finds
\begin{equation}
\hat \psi(z)\hat \psi(w)+\hat \psi(w)\hat \psi(z)=0.
\end{equation}
\indent In conclusion, we performed an explicit construction of the projections
of the microscopic operators onto the space of incompressible deformations
of the Laughlin state. As a main technical tool we used the exact large $N$ 
asymptotic for the norm of the IDLS. 
It is given by the dispersionless limit of the integrable Toda hierarchy.  
We explicitly built both negative and positive modes of the edge
density operator and obtained the central charge of their current algebra.  We
found that the fermion operators projected onto the low-energy subspace of the
space of IDLS have the properties predicted by the $\chi$LL theory. 
Thus, our main result is in establishing the one-to-one correspondence 
between the $\chi$LL and the space of incompressible deformations of the Laughlin
state. Understanding the experimental and numerical deviations from the $\chi$LL 
should be sought in non-trivial corrections to the Laughlin state, which in our 
formalism amounts to expanding the Hilbert space of the edge states and constructing 
the non-linear field theory in this bigger space. This work is planned for the
future.

The authors are grateful to P.B.Wiegmann for inspiring discussions and
collaboration at the early stage of this work. A.B. acknowledges financial
support of Swiss Science Foundation.

 \end{document}